# Introducing Time-dependent Molecular Fields: Derivation of the Wave Equations

by

Michael Baer,

The Fritz Haber Center for Molecular Dynamics, The Hebrew University of Jerusalem, Jerusalem 91904, Israel

---

[a]email: michaelb@fh.hujA.ac.il

# Abstract

This article is part of a series of articles trying to establish the concept *Molecular Field.* The theory that induced us to introduce this novel concept is based on the Born-Huang expansion as applied to the Schroedinger equation that describes the interaction of a molecular system with an external electric field. Assuming the molecular system is made up of *two* coupled adiabatic states the theory leads from a single spatial Curl Equation, two space-time Curl equations and one single space-time Divergent equation to a pair of decoupled Wave Equations usually encountered within the theory of fields. In the present study, just like in the previous one (see Baer, Mukherjee, Mukherjee and Adhikari, Molec. Phys., 114 , 227 (2016)) the Wave Equations are derived for an electric field having two features: (a) its intensity is *high enough*; (b) its duration is *short enough*. For this situation the study reveals that the just described interaction creates two fields that coexist within a molecule: one is a novel vectorial field formed via the interaction of the electric field with the Born-Huang non-adiabatic coupling terms (NACT) and the other is an ordinary, scalar, electric field essentially identical to the original electric field.

The present publication differs from the previous one in two ways: (1) The derivation of the Wave Equations is done in an entire different way which is more straightforward, fluent and easy to follow as compared to the previous one and therefore should be presented again for further studies. (2) In contrast to the previous publication the present results are analyzed via two intersecting Jahn-Teller cones which contain NACTs that become *singular* at the intersection point of these cones. A figure is given to show an electric beam, schematically imitated by wiggled red-green arrow, interacting with NACTs (in gray) "concentrated" inside the two intersecting conical potentials.

Finally, the fact that eventually we are applying mathematics associated with cosmic situations brings us to believe that (singular) NACTs are related to cosmic phenomena. Thus, if indeed this speculation is somehow connected to reality then, like other magnitudes in physics (expressed mathematically in terms of singularity points) the NACTs are formed at (or immediately after) the Big Bang and consequently, *guarantee* among many other physical magnitudes the formation of *molecular systems* during that event.

# I. Introduction

Most of the ideas I and my colleagues had about the interaction of an external field with molecules were published in a series of articles starting in 2003. Step by step, this series developed for more than a decade, testing new ideas until it reached its final publication just a year ago. [1-7,8a] The main difficulty we encountered was how to include time, rigorously, while carrying out the Born-Huang (BH) [9a] separation of variables. The straightforward way is to treat the field-dressed case – the time-dependent (TD) equation – just like one treats the field-free case, as originally proposed by Born and Oppenheimer (BO), [9b] namely by relating to the time-independent (TID) eigen-value equation: [1,3a]

$$\left(\mathbf{H}_{e0}(\mathbf{s_e},\mathbf{s}) - u_k(\mathbf{s})\right)|\zeta_k(\mathbf{s}_e|\mathbf{s})\rangle = 0; \quad k=\{1,N\} \tag{1}$$

Here $\mathbf{H}_{e0}(\mathbf{s_e},\mathbf{s})$ is the (field-free) electronic Hamiltonian, $u_k(\mathbf{s})$ are the corresponding electronic eigen-values (which are also recognized as the adiabatic potential energy surfaces (PES)), $|\zeta_k(\mathbf{s}_e|\mathbf{s})\rangle$; $k=\{1,N\}$ are the field-free (adiabatic) electronic eigen-functions and $\mathbf{s}$ and $\mathbf{s_e}$ stand for the

collection of *nuclear* and *electronic* coordinates respectively. In what follows this approach was termed as the "perturbative approach"[1]

The other possibility is to include time within the treatment of the relevant eigen-value problem. This change requires the inclusion of the corresponding time derivative so that Eq. (1) becomes: [1,2,3b]

$$i\hbar \frac{\partial}{\partial t}\left|\tilde{\zeta}_j(\mathbf{s}_e|\mathbf{s},t)\right\rangle = \mathbf{H}_e(\mathbf{s}_e,\mathbf{s},t)\left|\tilde{\zeta}_j(\mathbf{s}_e|\mathbf{s},t)\right\rangle; j=\{1,N\} \qquad (2)$$

Here $\mathbf{H}_e(\mathbf{s}_e,\mathbf{s},t)$ is the electronic Hamiltonian which also contains the external EM perturbation expressed, in the present article, in terms of an *electric* field interacting with the electrons, thus

$$\mathbf{H}_e(\mathbf{s}_e,\mathbf{s},t) = \mathbf{H}_{e0}(\mathbf{s}_e,\mathbf{s}) + \Phi(\mathbf{s}_e,\mathbf{s},t) \qquad (3a)$$

where

$$\Phi(\mathbf{s}_e,\mathbf{s},t) = \boldsymbol{\mu}_e(\mathbf{s}_e,\mathbf{s}) \cdot \boldsymbol{E}(t) \qquad (3b)$$

$\boldsymbol{\mu}_e(\mathbf{s}_e,\mathbf{s})$ is the electronic dipole moment defined as $\boldsymbol{\mu}_e(\mathbf{s}_e,\mathbf{s}) = e\sum \mathbf{s}_e(\mathbf{s})$ and $\boldsymbol{E}(t)$ is the intensity of the *electric* field. In what follows this approach was termed as the "non-perturbative approach"[1]

The advantage of the non-perturbative approach is in its physical message, namely it implies that the main effect of the external field is due to its interaction with the light, fast moving electrons and therefore it is expected that this situation has to be treated, from the onset, explicitly.

Numerically, the two approaches yield identical results (if no other approximations are done), as indeed was found in a detailed study of the photo-dissociation process of the $H_2^+$ cation: [6]

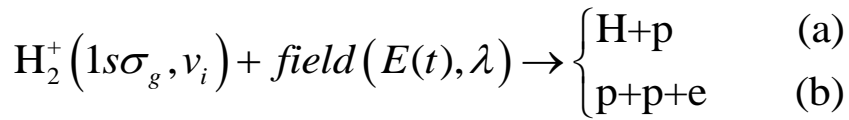

$$H_2^+(1s\sigma_g, v_i) + field(E(t), \lambda) \to \begin{cases} H+p & (a) \\ p+p+e & (b) \end{cases}$$

However from theoretical point of view the two approaches yield different results. Indeed, as was revealed already in two previous publications[7,8] and as will be discussed here the non-perturbative approach guarantees that the interaction of a polyatomic molecule with an external electric field leads to the concept of *Molecular Fields* which otherwise is missed. Such a possibility could eventually point at new processes taking place between molecules and an external electric field or even electro-magnetic (EM) fields.

At about the time when the perturbative and non-perturbative approaches were introduced we also re-considered the field-free *two-state* Curl equation [10,3c] and the corresponding, less known, *two-state* Divergent equation [11a,3d] (see Appendix A). While writing them down, we realized that they lead to the following (decoupled) Wave Equations: [11a,3e]

$$\frac{\partial^2 \tau_{z12}}{\partial p^2} + \frac{\partial^2 \tau_{z12}}{\partial q^2} = \frac{\partial}{\partial z'} \tau_{12}^{(2)}; \quad z = p, q; \ z' \neq z \quad (4)$$

where p and q are (mass dependent) nuclear coordinates so that each equation applies to one *single* spatial component of the NACTs, namely, $\tau_{p12}$, $\tau_{q12}$, etc. reminiscent of the Maxwell-Lorentz Wave Equation for the various components of the EM vector potential. [13] The above Wave Equations are expressed in terms of the two types of NACTs defined as:

$$\tau_{z12} = \left\langle \varsigma_1 \left| \frac{\partial}{\partial z} \varsigma_2 \right\rangle; \quad z = p, q \quad (5a)$$

and

$$\tau_{12}^{(2)} = \left\langle \varsigma_1 \left| \nabla^2 \varsigma_2 \right\rangle. \quad (5b)$$

and the indices 1 and 2 designate two adjacent electronic adiabatic states.

The main difference between the present Wave Equations and those encountered in the EM theory is that they are TID (namely, *stationary* Wave

Equations). This most straightforward result led us to think that eventually we may be able to form also a similar group of *time-dependent* Wave Equations.

Therefore the next step was to introduce time and this could be done if and only if the molecule was exposed to an external TD perturbation (viz. an external EM field). Since it was soon clear that employing the TID eigen-value equations (see Eq. (1)) would not lead to the expected TD Wave Equations we tried Eq. (2) and this turned out to be successful, as will be briefly discussed in the next Chapter.

## II. Background Information

In this section the aim is to enhance the connection between the theory to be presented here (see fractions in Refs. 7 and 8) with the approach which rests on the BH electronic NACTs and the fact that these magnitudes possess unique mathematical features somewhat strange to chemistry, namely, that they are singular (in fact poles) at certain points in configuration space of a given molecular system.

The BO[9b] approach to treat molecular systems is mainly known for introducing the concept of the *adiabatic* PESs, $u_j(\mathbf{s})$ j={1,….,N} and it was only later due to BH that attention was given to the above mentioned

NACTs; $\tau_{ij}(\mathbf{s})$; i,j = {1,….,N}, the coupling terms responsible for transitions between any pair of such adiabatic PESs. [9a] Here N is the size of the group of eigen-functions that form approximately a Hilbert space [3c] (see Appendix A). It is important to emphasize that N is dependent on the size of configuration space – the more extensive the configuration space the larger is N. Whereas the adiabatic PESs are well accepted and used routinely in numerical applications (although mainly in the case of the BO approximation), the NACTs are, in most numerical applications (of poly-atomic systems), ignored. From mathematical point of view the poly-atomic BH equations cannot be solved numerically unless the singularities, as produced by the Born-Oppenheimer-Huang (BOH) approach, are either "neutralized" in some way or eliminated. Rigorously speaking they can be eliminated by applying the *adiabatic-to-diabatic transformation* (ADT) [10] However, in numerous *quantum mechanical* applications the ADT is ignored and the diabatic PESs (in fact, all of quantum mechanical treatments are based on diabatic PESs) are reached by non-rigorous means. Thus it is difficult to accept that a theory based on two equally important components may yield meaningful numerical results when one of the two is ignored. In the Reference section are listed a series of publications which to some extent enable the less knowledgeable reader to get acquainted with the literature on the NACTs (see refs. 1-8, 10-12, 14-59). In particular I refer to studies by Englman and Vertesi [46] who, employing Berry's adiabatic Phase, analyzed some of the topological phases calculated by Baer et al. [30] enhancing the connection between the two approaches. Also, I would like to mention a recent interesting approach, named Factorization, by Gross et al. [47,48] who treat the BOH approach, while replacing the BH expansion by a single BO-like term (obviously − interpreted differently).

## III. Wave Equations for the Field Dressed Case

Although the Wave Equations were derived sometime ago,[7,8] here we concentrate on certain aspects related to this derivation. The previous treatment was complicated and demanded a lot of algebra. Also due to uncertainties we devoted efforts inquiring various routes to guarantee the relevance of the algebraic manipulations etc. Moreover various models were treated as part of the derivation to make sure the assumptions made during this process are meaningful and reliable. As an <u>example</u> I refer to the two-state TID Curl equation and the corresponding TD Curl equation. In the first case the NACT matrix contains only two terms that differ from zero whereas in the second, TD case, all four elements differ from zero. Therefore there is place to worry if features that are valid in case of the TID Curl equation apply also for the TD Curl equation. Finally, since the resultant Wave Equations were already known we could devise a more efficient way to derive them as, indeed, is presented here. Thus, whereas the treatments in Refs. (7) and (8) yield complicated expressions the derivation in the present publication is straightforward and easy-to-follow and therefore it is important to expose it.

In the previous articles [7,8] we distinguished between the treatment of the left hand side (l.h.s,) of these equations and their right hand side (r.h.s.). The reason being that the l.h.s. is based solely on the l.h.s. of the mathematical operators, namely, the Curl Equations and the Divergence (Eqs. 26-29 and (32) - see below). The derivation of the r.h.s. depends on both, the r.h.s. of the mathematical operators and the explicit kind of field interacting with the molecule as expressed by the {Field-Free →Field Dressed} Transformation Matrix $\omega(\mathbf{s},t)$ (see Appendix B). In the present study the $\omega(\mathbf{s},t)$-matrix is derived for the case that the external field is an *electric* field characterized by a high intensity (see Eq. (B.11)) enable to ignore certain terms and short-time duration (as compared to nuclear motion) enable to ignore the derivatives of the external field with respect to the nuclear coordinates p and q (see Appendix C). It is important to emphasize that any assumptions related to the external field do not affect the *structure* of the Wave Equations as is well noticed from Eqs. (37a), (37b) and (38) (see below).

## III.1. Deriving the Adiabatic Schrödinger Equation

The nuclear-electron Schrödinger Equation to be treated is of the form:

$$i\hbar \frac{\partial \mathbf{\Psi}(\mathbf{s}_e,\mathbf{s},t)}{\partial t} = \left(-\frac{\hbar^2}{2m}\nabla^2 + \mathbf{H}_e(\mathbf{s}_e,\mathbf{s},t)\right)\mathbf{\Psi}(\mathbf{s}_e,\mathbf{s},t) \tag{6}$$

where m is the mass of the (nuclear) system, $\nabla$ is the gradient (vector) operator expressed in terms of mass-scaled coordinates and $\mathbf{H}_e(\mathbf{s}_e,\mathbf{s},t)$ is the electronic Hamiltonian which becomes TD only for t≥0.

Next is applied the BH expansion:

$$\mathbf{\Psi}(\mathbf{s}_e,\mathbf{s},t) = \tilde{\boldsymbol{\zeta}}^T(\mathbf{s}_e|\mathbf{s},t)\tilde{\boldsymbol{\psi}}(\mathbf{s},t) \tag{7}$$

where $\tilde{\boldsymbol{\zeta}}^T(\mathbf{s}_e|t,\mathbf{s})$ is a row vector. Substituting Eqs. (2) and (7) in Eq.(6) yields the following somewhat unexpected expression:

$$i\hbar\tilde{\boldsymbol{\zeta}}^T(\mathbf{s}_e|\mathbf{s},t)\frac{\partial \tilde{\boldsymbol{\psi}}(\mathbf{s},t)}{\partial t} = -\frac{\hbar^2}{2m}\nabla^2\left(\tilde{\boldsymbol{\zeta}}^T(\mathbf{s}_e|\mathbf{s},t)\tilde{\boldsymbol{\psi}}(\mathbf{s},t)\right) \tag{8}$$

The interesting feature of this equation is its simplistic form – it just contains the two functions that form the BH expression (see Eq. (7)) but none of the magnitudes that define the system, in particular, the PESs. Evaluating the derivatives on the r.h.s. yields:

$$i\hbar\tilde{\zeta}^T \frac{\partial \tilde{\psi}}{\partial t} = -\frac{\hbar^2}{2m}\{\tilde{\zeta}^T \nabla^2 \tilde{\psi} + 2(\nabla \tilde{\zeta}^T)\cdot(\nabla \tilde{\psi}) + (\nabla^2 \tilde{\zeta}^T)\tilde{\psi}\} \quad (9)$$

Multiplying the resulting equation (i.e. Eq. (9)) by the column vector $\tilde{\zeta}(\mathbf{s}_e | t, \mathbf{s})$ and integrating over electronic coordinates yields:

$$i\hbar \frac{\partial}{\partial t}\tilde{\psi}_k = -\frac{\hbar^2}{2m}\nabla^2 \tilde{\psi}_k - \frac{\hbar^2}{2m}\sum_{j=1}^{N}\left(2\tilde{\tau}_{kj}\cdot\nabla + \tilde{\tau}^{(2)}_{kj}\right)\tilde{\psi}_j = 0; \quad k=\{1,N\} \quad (10a)$$

where

$$\tilde{\tau}_{jk} = \langle \tilde{\zeta}_j | \nabla \tilde{\zeta}_k \rangle \quad (11a)$$

and

$$\tilde{\tau}^{(2)}_{jk} = \langle \tilde{\zeta}_j | \nabla^2 \tilde{\zeta}_k \rangle \quad (11b)$$

are the ordinary and the second order *field dressed* (TD) NACTs respectively.

The connection between $\mathbf{\tau(s)}$ and $\tilde{\mathbf{\tau}}(\mathbf{s},t)$ is given next

$$\tilde{\mathbf{\tau}}(\mathbf{s},t) = \mathbf{\omega}^\dagger \mathbf{\tau(s)}\mathbf{\omega} + \mathbf{\omega}^\dagger \nabla \mathbf{\omega} \quad (11c)$$

as was derived for the first time in Ref. 1 and then, further discussed, in Refs. (2) and (3b). We recall that $\tau(\mathbf{s})$ is the *Field-Free* NACT matrix (see Eq. (5a)) and $\boldsymbol{\omega}(\mathbf{s},t)$ is the {Field-Free $\to$ Field-Dressed} transformation matrix (see Appendix B).

Eq. (10a) can be written in a more compact way, similar to its form in the field-free case: [1,3b]

$$i\hbar \frac{\partial \tilde{\psi}}{\partial t} = -\frac{\hbar^2}{2m}\left(\nabla + \tilde{\boldsymbol{\tau}}\right)^2 \tilde{\psi} \tag{10b}$$

Eq. (10b) completes the derivation of the field-dressed adiabatic BH equation. Eq. (10b) lacks the adiabatic potential matrix because such a potential does not exist within the TD framework.

## III.2. The Adiabatic-to-Diabatic Transformation matrix, $\tilde{\mathbf{A}}$, in Space-Time Configurations

In order to expose the Curl equation required for constructing the Wave Equations we have to discuss briefly an issue not directly associated with these equations. Eq. (10b) cannot be solved numerically as such because some of the NACT elements, $\tilde{\tau}_{jk}$, are expected to be singular in configuration space. In order to overcome this difficulty these matrix elements have to be eliminated and this can be done by the earlier mentioned ADT carried out employing a matrix $\tilde{\mathbf{A}}(\mathbf{s},t)$. The elements of this matrix have to be analytic functions of the nuclear coordinates, a requirement that will be shown to lead to the various Curl equations. We intend to refer to this issue only briefly as it was discussed frequently in previous publications. [1,2, 3g,7]

The elimination of the NACTs will be done by replacing, in Eq. (10b), the column $\tilde{\boldsymbol{\psi}}(\mathbf{s},t)$ with the product $\tilde{\mathbf{A}}(\mathbf{s},t)\tilde{\boldsymbol{\chi}}(\mathbf{s},t)$, thus:

$$i\hbar\tilde{\mathbf{A}}\frac{\partial\tilde{\boldsymbol{\chi}}}{\partial t} = -\frac{\hbar^2}{2m}\left[\tilde{\mathbf{A}}\nabla^2 + 2(\mathbf{G}\tilde{\mathbf{A}})\nabla + (\mathbf{G}^2\tilde{\mathbf{A}})\right]\tilde{\boldsymbol{\chi}} - i\hbar\frac{\partial\tilde{\mathbf{A}}}{\partial t}\tilde{\boldsymbol{\chi}} \qquad (12)$$

Here, $\mathbf{G}(\mathbf{s},t)$ is the space *covariant derivative*.[7,3g] operator which acts only on $\tilde{\mathbf{A}}(\mathbf{s},t)$ and is given in the form:

$$\mathbf{G} = \nabla + \tilde{\boldsymbol{\tau}} \qquad (13)$$

To continue, we add and subtract an undetermined matrix $\boldsymbol{\tau}_t(\mathbf{s},t)$ multiplied by $\tilde{\mathbf{A}}(\mathbf{s},t)\tilde{\boldsymbol{\chi}}(\mathbf{s},t)$, so that Eq. (12) becomes:

$$i\hbar\tilde{\mathbf{A}}\frac{\partial\tilde{\boldsymbol{\chi}}}{\partial t} = -\frac{\hbar^2}{2m}\left[\tilde{\mathbf{A}}\nabla^2 + 2(\mathbf{G}\tilde{\mathbf{A}})\nabla + (\mathbf{G}^2\tilde{\mathbf{A}})\right]\tilde{\boldsymbol{\chi}} - (\mathbf{G}_t\tilde{\mathbf{A}} - \tilde{\boldsymbol{\tau}}_t\tilde{\mathbf{A}})\tilde{\boldsymbol{\chi}} \qquad (14)$$

where $\mathbf{G}_t(\mathbf{s},t)$ is the corresponding *time covariant derivative*.[7] operator defined as:

$$\mathbf{G}_t = i\hbar \frac{\partial}{\partial t} + \tilde{\boldsymbol{\tau}}_t \qquad (15)$$

and $\tilde{\boldsymbol{\tau}}_t$ is assumed to be an Hermitian operator (a fact that is verified elsewhere [1,3g]).

To continue, we require that the $\tilde{\mathbf{A}}$-matrix has to fulfill the following spatial first order differential equations: [1,2,3g,7]

$$\mathbf{G}\tilde{\mathbf{A}} = \mathbf{0} \quad \Rightarrow \quad \nabla\tilde{\mathbf{A}} + \tilde{\boldsymbol{\tau}}\tilde{\mathbf{A}} = \mathbf{0} \qquad (16a)$$

as well as the corresponding temporal first order differential equation

$$\mathbf{G}_t\tilde{\mathbf{A}} = \mathbf{0} \quad \Rightarrow \quad i\hbar\frac{\partial \tilde{\mathbf{A}}}{\partial t} + \tilde{\boldsymbol{\tau}}_t\tilde{\mathbf{A}} = \mathbf{0} \qquad (16b)$$

Substituting Eqs. (16) in Eq. (14) yields the corresponding field dressed *diabatic* Schrödinger equation:

$$i\hbar \frac{\partial \tilde{\boldsymbol{\chi}}(\mathbf{s},t)}{\partial t} = \left(-\frac{\hbar^2}{2m}\nabla^2 + \tilde{\mathbf{W}}_e(\mathbf{s},t)\right)\tilde{\boldsymbol{\chi}}(\mathbf{s},t) \qquad (17)$$

where $\tilde{\boldsymbol{\chi}}(\mathbf{s},t)$ is recognized as a column vector that contains the *diabatic* nuclear wave functions and $\tilde{\mathbf{W}}_e(\mathbf{s},t)$ is the corresponding field-dressed *diabatic* PES matrix that follows from the equation

$$\tilde{W}_e(s,t) = \tilde{A}(s,t)^\dagger \tilde{\tau}_t(s,t) \tilde{A}(s,t) \qquad (18)$$

More details regarding the ADT are given elsewhere.[1,3g,7]

**Comment**: The derivation presented so far is not complete unless the matrix $\tilde{\tau}_t(s,t)$ is explicitly specified. This matrix takes the form:[1,3g]

$$\tilde{\tau}_t = \omega^\dagger \tilde{H}_e \omega \qquad (19)$$

where $\omega(s,t)$, which contains the {Field Free $\to$ Field Dressed} *expansion* coefficients, is given in the form:

$$\omega(s,t) = \exp\left\{ -\frac{i}{\hbar} \int_0^t \tilde{H}_e(s,t') dt' \right\}, \qquad (20)$$

In Appendix B the $\omega(s,t)$-matrix is derived for the *high intensity* limit, namely

$$\tilde{\Phi}_{e12} \gg \frac{1}{2}\left|(\tilde{u}_1 - \tilde{u}_2)\right| \qquad (21)$$

where $\tilde{u}_j(s,t)$; j=1,2 are defined in Eq. (B8) (see also Eq. (B2)).

Consequently $\omega(s,t)$ was found to be of the form:

$$\boldsymbol{\omega}(\mathbf{s},t) = \begin{pmatrix} e^{i\theta_1} & 0 \\ 0 & e^{-i\theta_2} \end{pmatrix} \begin{pmatrix} \cos(\theta/2) & \sin(\theta/2) \\ -\sin(\theta/2) & \cos(\theta/2) \end{pmatrix}$$

or:

$$\boldsymbol{\omega}(\mathbf{s},t) = \begin{pmatrix} e^{i\theta_1}\cos(\theta/2) & e^{i\theta_1}\sin(\theta/2) \\ -e^{-i\theta_2}\sin(\theta/2) & e^{-i\theta_2}\cos(\theta/2) \end{pmatrix}$$

(22)

where the phases $\theta_1(\mathbf{s},t)$ and $\theta_2(\mathbf{s},t)$ are explicitly presented there (see Eqs (B.21)) thus

$$\theta_1(\mathbf{s},t) = -\frac{1}{\hbar}\int_0^t \left\{ \frac{1}{2}\left[\tilde{u}_1(\mathbf{s},t') + \tilde{u}_2(\mathbf{s},t')\right] + \tilde{\Phi}_{e12}(\mathbf{s},t') \right\} dt'$$

$$\theta_2(\mathbf{s},t) = \frac{1}{\hbar}\int_0^t \left\{ \frac{1}{2}\left[\tilde{u}_1(\mathbf{s},t') + \tilde{u}_2(\mathbf{s},t')\right] - \tilde{\Phi}_{e12}(\mathbf{s},t') \right\} dt'$$

(23)

as well as the phase $\theta(\mathbf{s},t)$

$$\theta(\mathbf{s},t) = \tan^{-1}\left\{ \frac{\tilde{\Phi}_{e12}(\mathbf{s},t) - [\tilde{u}_2(\mathbf{s},t) - \tilde{u}_1(\mathbf{s},t)]}{\tilde{\Phi}_{e12}(\mathbf{s},t)} \right\}$$

(24a)

In addition we also frequently refer to the phase $\Theta(\mathbf{s},t)$:

$$\Theta(\mathbf{s},t) = \theta_1(\mathbf{s},t) + \theta_2(\mathbf{s},t) = -\frac{2}{\hbar}\int_0^t \tilde{\Phi}_{e12}(\mathbf{s},t')dt' \qquad (25)$$

which stands for the sum of the two phases.

For certain purposes the approximation in Eq. (21) can be moved one step further, namely, applying it in Eq. (24) so that θ(**s**,*t*) essentially becomes a constant:

$$\theta(\mathbf{s},t) = \pi/2 \Rightarrow \cos\frac{\theta}{2} = \sin\frac{\theta}{2} = \sqrt{2}/2 \text{ and } \cos\theta = 0;\ \sin\theta = 1$$

(24b)

**Comment(1)**: Whenever we refer to ω(**s**,*t*)-matrix, either in the main body of the article or in Appendix B, it will be the matrix given in Eq. (22).

**Comment(2)**: It is important to emphasize that the present choice of parameters was made to guarantee *easy to derive and more transparent results*.

## III.3. Curl and Divergent equations

### III.3.1 Curl Equations

The origin of the Curl equations is associated with the requirement that the diabatic potential matrix, $\tilde{\mathbf{W}}_e(\mathbf{s},t)$, just like any other potential, has to be single-valued, namely, analytic.[10] From Eq. (17b) it is noticed that

$\tilde{\mathbf{W}}_e(\mathbf{s},t)$ is dependent on the ADT matrix, $\tilde{\mathbf{A}}(\mathbf{s},t)$, and in order for its elements to be an analytic function of the coordinates, this matrix has to be analytic with regard to the nuclear coordinates $\mathbf{s}=\{p,q,\ldots\}$ and with regard to time, t.

In Refs (7) and (8) are discussed the following (matrix) Curl equations:

(i) The spatial (p,q) component of the Curl equation

$$\tilde{\mathbf{H}}_{pq} = \tilde{\mathbf{T}}_{pq} \tag{26}$$

where $\tilde{\mathbf{H}}_{pq}$ and $\tilde{\mathbf{T}}_{pq}$ are matrices defined in terms of $\tilde{\boldsymbol{\tau}}_p$ and $\tilde{\boldsymbol{\tau}}_q$:

$$\begin{aligned}\tilde{\mathbf{H}}_{pq} &= \frac{\partial \tilde{\boldsymbol{\tau}}_p}{\partial q} - \frac{\partial \tilde{\boldsymbol{\tau}}_q}{\partial p} \quad &\textbf{(a)}\\ \tilde{\mathbf{T}}_{pq} &= \tilde{\boldsymbol{\tau}}_p \tilde{\boldsymbol{\tau}}_q - \tilde{\boldsymbol{\tau}}_q \tilde{\boldsymbol{\tau}}_p \quad &\textbf{(b)}\end{aligned} \tag{27}$$

(ii) The space-time (z,t) components of the Curl equation:

$$\tilde{\mathbf{H}}_{zt} = \tilde{\mathbf{T}}_{zt}; \quad z=p,q \tag{28}$$

where $\tilde{\mathbf{H}}_{zt}$ and $\tilde{\mathbf{T}}_{zt}$ are matrices defined in terms of $\tilde{\boldsymbol{\tau}}_t$ and $\tilde{\boldsymbol{\tau}}_z$:

$$\begin{aligned}\tilde{\mathbf{H}}_{zt} &= i\hbar\frac{\partial \tilde{\boldsymbol{\tau}}_z}{\partial t} - \frac{\partial \tilde{\boldsymbol{\tau}}_t}{\partial z}; \quad &z=p,q \;\; (a)\\ \tilde{\mathbf{T}}_{zt} &= \tilde{\boldsymbol{\tau}}_z \tilde{\boldsymbol{\tau}}_t - \tilde{\boldsymbol{\tau}}_t \tilde{\boldsymbol{\tau}}_z; \quad &z=p,q \;\; (b)\end{aligned} \tag{29}$$

These Curl equations are used to construct the Wave Equations for certain field dressed NACTs *elements* (not the matrices themselves) as will be done in Section III.3.3.

### III.3.2 Divergent Equation

The space-time Divergent $(\nabla\tilde{\boldsymbol{\tau}})_{pqt}$ is formed by a linear combination of the spatial Divergence $(\nabla\tilde{\boldsymbol{\tau}})_{pq}$ given (in case of two coordinates p and q) in the form:

$$(\nabla\tilde{\boldsymbol{\tau}})_{pq} = \frac{\partial\tilde{\boldsymbol{\tau}}_p}{\partial p} + \frac{\partial\tilde{\boldsymbol{\tau}}_q}{\partial q} \tag{30}$$

and the corresponding time derivative of $\tilde{\boldsymbol{\tau}}_t$ (see Eqs. (19)), i.e., $\partial\tilde{\boldsymbol{\tau}}_t/\partial t$. In reference (7) it was suggested to extend the *Field-Free* divergent equation as follows:

$$(\nabla\tilde{\boldsymbol{\tau}})_{pqt} = \frac{\partial\tilde{\boldsymbol{\tau}}_p}{\partial p} + \frac{\partial\tilde{\boldsymbol{\tau}}_q}{\partial q} + i\kappa\hbar\frac{\partial\tilde{\boldsymbol{\tau}}_t}{\partial t} \tag{31}$$

were κ is a parameter which ensures that the dimensions of the first two terms are consistent with the dimensions of the third term. At this stage we just mention that $\kappa = (\hbar v)^{-2}$ where v is the group velocity associated with the *Molecular Field*. Consequently Eq. (31) becomes:

$$(\nabla \tilde{\boldsymbol{\tau}})_{pqt} = \frac{\partial \tilde{\boldsymbol{\tau}}_p}{\partial p} + \frac{\partial \tilde{\boldsymbol{\tau}}_q}{\partial q} + i\frac{1}{\hbar v^2}\frac{\partial \tilde{\boldsymbol{\tau}}_t}{\partial t} \tag{32}$$

This way of forming the divergence equation is reminiscent – but still different – of how it is derived within the EM theory.[13]

## III.3.3 Preparing the Curl and Divergence Equations for the Wave-Equations

Before starting the analysis to be given next, we emphasize that the study applies for 2x2 matrices only.

In order to form the Wave Equations we need to extract from Eqs. (27), (29) and (32) a *single characteristic matrix element* which applies for both the Field-Free case (the stationary case) and the corresponding Field-Dressed case (i.e., when the molecular system is exposed to an external field). In Appendix I it is shown that for deriving the stationary Wave Equation we have to form the relevant Curl equation and Divergent equation for the *off-diagonal* (1,2) matrix element which is responsible for the Abelian case. The *continuity* requirement dictates the choice of the relevant matrix elements also for the Field-Dressed case, namely the corresponding (1,2) matrix element (although the matrices are not Abelian anymore).

### III.3.3.1 Deriving the Curl and Divergent equations for matrix elements

Due to Eqs. (26) and (27) (and various remarks), the relevant spatial Curl equation for the corresponding (1,2) *matrix elements* are:

$$\frac{\partial \tilde{\tau}_{p12}}{\partial q} - \frac{\partial \tilde{\tau}_{q12}}{\partial p} = \left( \tilde{\tau}_p \tilde{\tau}_q - \tilde{\tau}_q \tilde{\tau}_p \right)_{12} = C_{pq}(p,q,t) \qquad (33)$$

In the same way the relevant space-time Curl equations for the corresponding (1,2) *matrix elements* that follow from Eqs. (28) and (29) are:

$$i\hbar \frac{\partial \tilde{\tau}_{q12}}{\partial t} - \frac{\partial \tilde{\tau}_{t12}}{\partial q} = \left( \tilde{\tau}_q \tilde{\tau}_t - \tilde{\tau}_t \tilde{\tau}_q \right)_{12} = C_{qt}(p,q,t) \qquad (34)$$

and

$$i\hbar \frac{\partial \tilde{\tau}_{p12}}{\partial t} - \frac{\partial \tilde{\tau}_{t12}}{\partial p} = \left( \tilde{\tau}_p \tilde{\tau}_t - \tilde{\tau}_t \tilde{\tau}_p \right)_{12} = C_{pt}(p,q,t) \qquad (35)$$

In the same way the relevant (1,2) Divergent equation (extracted from Eq. (32)) takes the form :

$$\frac{\partial \tilde{\tau}_{p12}}{\partial p} + \frac{\partial \tilde{\tau}_{q12}}{\partial q} + i \frac{1}{\hbar v^2} \frac{\partial \tilde{\tau}_{t12}}{\partial t} = D_{pqt}(p,q,t) \qquad (36)$$

## III.4 Wave Equations

### III.4.1 Left-Hand-Side of the Wave Equations

To form the Wave Equation for $\tilde{\tau}_{p12}$ we form the following expression:

$$\therefore \quad \frac{\partial}{\partial q}(Eq.33) + \frac{\partial}{\partial p}(Eq.36) + i\frac{1}{v^2\hbar}\frac{\partial}{\partial t}(Eq.35)$$

and obtain the result:

$$\frac{\partial^2 \tilde{\tau}_{p12}}{\partial q^2} + \frac{\partial^2 \tilde{\tau}_{p12}}{\partial p^2} - \frac{1}{v^2}\frac{\partial^2 \tilde{\tau}_{p12}}{\partial t^2} = J_p(p,q,t) \tag{37a}$$

A similar procedure is carried out for Wave Equation of $\tilde{\tau}_{q12}$, namely forming the following expression:

$$\therefore \quad -\frac{\partial}{\partial p}(Eq.33) + \frac{\partial}{\partial q}(Eq.36) + i\frac{1}{v^2\hbar}\frac{\partial}{\partial t}(Eq.34)$$

and obtain the result:

$$\frac{\partial^2 \tilde{\tau}_{q12}}{\partial p^2} + \frac{\partial^2 \tilde{\tau}_{q12}}{\partial q^2} - \frac{1}{v^2}\frac{\partial^2 \tilde{\tau}_{q12}}{\partial t^2} = J_q(p,q,t) \tag{37b}$$

and finally, to form the Wave Equation for $\tilde{\tau}_{t12}$ we form a similar expression:

$$\therefore \quad -\frac{\partial}{\partial q}(Eq.34) - \frac{\partial}{\partial p}(Eq.35) + i\hbar\frac{\partial}{\partial t}(Eq.36)$$

so that the corresponding Wave Equation becomes:

$$\frac{\partial^2 \tilde{\tau}_{t12}}{\partial p^2} + \frac{\partial^2 \tilde{\tau}_{t12}}{\partial q^2} - \frac{1}{v^2}\frac{\partial^2 \tilde{\tau}_{t12}}{\partial t^2} = \rho(p,q,t) \tag{38}$$

It is noticed that our treatment so far, yields two kinds of Wave Equations: the *decoupled* equations in Eqs.(37) which apply for the p and q components of the Field-Dressed (vectorial) NACTs and a third equation in Eq. (38) for the $\tilde{\tau}_{t12}$-matrix element.

## III.4.2 Right-Hand-Side of the Wave Equations

The explicit expressions for the r.h.s. follow from algebraic manipulations and differentiations performed with respect to NACT elements that constitute the matrices $\tilde{\boldsymbol{\tau}}_z(\mathbf{s},t);\ z = p,q$ and $\tilde{\boldsymbol{\tau}}_t(\mathbf{s},t)$.

In order to carry out these calculations we need the explicit expressions for three (2x2) matrices, namely, the matrix $\tilde{\boldsymbol{\tau}}_t(\mathbf{s},t)$ (see (19)) and the two matrices $\tilde{\boldsymbol{\tau}}_z(\mathbf{s},t);\ z = p,q$ which take the form (see Eqs. (11c)):

$$\tilde{\boldsymbol{\tau}}_z(\mathbf{s},t) = \boldsymbol{\omega}^\dagger \boldsymbol{\tau}_z(\mathbf{s})\boldsymbol{\omega} + \boldsymbol{\omega}^\dagger \frac{\partial \boldsymbol{\omega}}{\partial z} \tag{39}$$

Here $\boldsymbol{\tau}_z$ (s); z=p,q are the *Field-Free* NACT matrices (see Eq. (5a)) and

$\boldsymbol{\omega}(s,t)$ is the Field-Free → Field-Dressed transformation matrix

(see Eq. 22). [1,2]

In Appendix B of Ref. 8 are derived the explicit expressions for

$\tilde{\boldsymbol{\tau}}_z(s,t)$; z=p,q:

$$\tilde{\boldsymbol{\tau}}_z = \begin{pmatrix} i\left(\tau_{z12}\sin\Theta\sin\theta + \cos^2\left(\frac{\theta}{2}\right)\frac{\partial\theta_1}{\partial z} - \sin^2\left(\frac{\theta}{2}\right)\frac{\partial\theta_2}{\partial z}\right) & (\cos\Theta - i\cos\theta\sin\Theta)\tau_{z12} + \frac{1}{2}\left(i\sin\theta\frac{\partial\Theta}{\partial z} + \frac{\partial\theta}{\partial z}\right) \\ -(\cos\Theta + i\cos\theta\sin\Theta)\tau_{z12} + \frac{1}{2}\left(i\sin\theta\frac{\partial\Theta}{\partial z} - \frac{\partial\theta}{\partial z}\right) & -i\left(\tau_{z12}\sin\Theta\sin\theta - \sin^2\left(\frac{\theta}{2}\right)\frac{\partial\theta_1}{\partial z} + \cos^2\left(\frac{\theta}{2}\right)\frac{\partial\theta_2}{\partial z}\right) \end{pmatrix}$$
(40)

as well as for $\tilde{\boldsymbol{\tau}}_t(s,t)$

$$\tilde{\boldsymbol{\tau}}_t = -\hbar \begin{pmatrix} \cos^2(\frac{\theta}{2})\frac{\partial\theta_1}{\partial t} - \sin^2(\frac{\theta}{2})\frac{\partial\theta_2}{\partial t} & \frac{1}{2}\left(\sin\theta\frac{\partial\Theta}{\partial t} - i\frac{\partial\theta}{\partial t}\right) \\ \frac{1}{2}\left(\sin\theta\frac{\partial\Theta}{\partial t} + i\frac{\partial\theta}{\partial t}\right) & \sin^2(\frac{\theta}{2})\frac{\partial\theta_1}{\partial t} - \cos^2(\frac{\theta}{2})\frac{\partial\theta_2}{\partial t} \end{pmatrix} \quad (41)$$

The rest of the theoretical treatment is carried out only for the (upper) *off-diagonal* terms of the above three matrices. The present study is limited for the case the intensity $\tilde{\Phi}_{e12}$, due to the external field, is high enough (see Eqs. (21)) and its duration is short enough *for the nuclei barely to move.* [8]
Consequently these matrix elements acquire a simplified form (see Appendix C):

$$\tilde{\tau}_{z12}(\mathbf{s},t) \sim \cos\Theta(\mathbf{s},t)\tau_{z12}(\mathbf{s}); \quad z=p,q \tag{42}$$

and

$$\tilde{\tau}_{t12} \sim -\hbar\frac{1}{2}\frac{\partial\Theta}{\partial t} \tag{43}$$

In Appendix C are derived, for this situation, the explicit expressions of $C_{pq}$, $C_{pt}$, $C_{qt}$ and $D_{pqt}$ ( see Eqs. (33)-(36)) as finally given in Eqs.(C.1b'), (C.2b), (C.3b) and (C.4b') respectively:

$$C_{pq} = 0 \tag{33'}$$

$$C_{pt} = -i\hbar\sin\Theta\,\tau_{p12}(p,q)\frac{\partial\Theta}{\partial t} \tag{34'}$$

$$C_{qt} = -i\hbar\sin\Theta\,\tau_{q12}(p,q)\frac{\partial\Theta}{\partial t} \tag{35'}$$

$$D_{pqt} = \cos\Theta \, \tau_{12}^{(2)}(p,q) - \frac{i}{2}\frac{1}{c^2}\frac{\partial^2 \Theta}{\partial t^2} \tag{36'}$$

To form the inhomogeneities, $J_z(p,q,t)$, (see Eqs. (37a) and (37b)) the following two differentiations have to be carried out and added up:

$$\therefore \frac{\partial}{\partial z}(Eq.36') + i\frac{1}{v^2\hbar}\frac{\partial}{\partial t}(Eq.35')(or \frac{\partial}{\partial t}(Eq.34')) \tag{44}$$

(where due to the additional assumption: $\partial\Theta/\partial z \equiv 0$ – see Appendix C )
Thus, the r.h.s. of Eqs. (37a) and (37b) can be shown to be:

$$J_z(p,q,t) = \cos\Theta \frac{\partial \tau_{12}^{(2)}}{\partial z} - \frac{1}{c^2}\tau_{z12}\frac{\partial^2(\cos\Theta)}{\partial t^2}; \quad z=p,q$$

$$\tag{45}$$

where

$$\Theta(p,q,t) = -\frac{2}{\hbar}\int_0^t \tilde{\Phi}_{e12}(p,q,t')dt' \tag{46}$$

(see Eq. (25)) and $\tau_{12}^{(2)}$ and $\tau_{z12}$ are introduced in Eqs. (4) and (5). In the same way the r.h.s.. of Eq.(38), namely, $\rho(p,q,t)$ follows from the expression in Eq. (47):

$$\therefore \frac{\partial}{\partial t}(Eq.36') - i\frac{1}{v^2\hbar}\frac{\partial}{\partial p}(Eq.34') - i\frac{1}{v^2\hbar}\frac{\partial}{\partial q}(Eq.35') \qquad (47)$$

which can be shown to yield

$$\rho(p,q,t) = \frac{1}{c^2}\frac{\partial^2 \tilde{\Phi}_{e12}(p,q,t)}{\partial t^2} \qquad (48)$$

It could be interesting to mention that the same results were obtained in Ref, (8) following an entirely different and much more complicated derivation.

It is important to emphasize that the l.h.s. of the Wave Equations are general and apply for any set of matrix elements: $\tilde{\tau}_{pjk}(s,t)$, $\tilde{\tau}_{qjk}(s,t)$, $\tilde{\tau}_{tjk}(s,t)$ but the r.h.s. depends on the kind of external field to be studied.

## III.5 Short Summary

To conclude this part we say that following the theory given here we managed to prove that, indeed, the interaction of a polyatomic molecule with an external *electric* produces Wave Equations which like in the EM case lead to Molecular Fields field. Our confidence is based on the fact that according to the theory, two types of fields are found to *coexist* within a molecule interacting with an external electric field: one is the novel

Molecular field, produced via Vector Potentials and the other is just the original, essentially unaffected, external electric field described by a scalar Wave Equation.

While scanning through the three decoupled Wave Equations, given in Eqs, (37a), (37b) and Eq. (38) and compare them with the corresponding EM Maxwell-Lorentz equations the following is noticed: The equations for the components of the Field-Dressed NACT are essentially similar to those for the components of the EM *vector potential* and the equation for the unknown function, $\tilde{\tau}_{t12}$, compares nicely with the equation for the scalar electric potential which is formed by the electric charge.

Another issue in this respect is as follows: Could it be that other approaches, although being gauge invariant, may lead to different findings? Our approach is based on the assumption that the external EM field interacts primarily with the *electrons* and this fact is incorporated by solving an electronic TD eigen-value problem (see Eq. (2)) instead of the usual, TID eigen-value equation (see Eq. (1)). By doing it this way we claim that our findings are germane and trustworthy. But there are the other treatments (a sample is listed below – see Refs. 60 -78) based on Eq. (1) and therefore do not expose the existence of the Molecular Field. In other words, the fact that the Molecular Field was not revealed earlier is a mishap due to the less transparent treatment of the electronic eigenvalue issue.

Finally we briefly mention a somewhat related subject, namely, the occurrence of light-induced conical intersections in diatomic molecules exposed to light fields. [79] This situation can be relevant to the present study if the emerged perturbed light field is affected by the field-free NACTs.

## IV. Discussion – Introducing the Jahn-Teller Model

At this stage I speculate on the meaning of this novel field and for this purpose I apply the Jahn-Teller (JT) model which is known to be made up of two potentials that take the form of two inverted cones, intersecting at a point (see Fig. 1). The two potentials in polar coordinates, i.e., $u_1(\varphi,r)$ and $u_2(\varphi,r)$, are of the form:

$$u_1(\varphi,r) = kr \tag{49a}$$

$$u_2(\varphi,r) = -kr \tag{49b}$$

where $0 \leq \varphi \leq 2\pi$. The point of intersection is known as the point of the conical intersection or as the *ci*. [3c,80-81]

For this model it can be shown [3h] that the two relevant (planar) NACTs are:

$$\tau_\varphi(\varphi,r) = -\frac{1}{2r}$$
$$\tau_r(\varphi,r) = 0 \qquad (50)$$

From this presentation it is noticed that, $\tau_\varphi(\varphi,q)$, the angular NACT, gets larger the closer the relevant configuration space is to the *ci* – the point where $\tau_\varphi$ becomes infinite

In Fig. 1 is also shown, schematically, an electric beam that interacts with the NACTs inside the cones and eventually produces the Molecular Field. In Eq. (45) are presented the in-homogeneities of the resulting Wave Equations which are seen to intensify the larger are the *field-free* NACTs, namely, $\tau_{z12}$; z=p,q. The JT model is expressed in terms of polar coordinates ($\varphi$,q) and therefore the NACT to get intensified in this case is only $\tau_\varphi(\varphi,q)$ and due to Eq. (50) this happens the closer the beam gets to the ci-point (see Fig. 1). Indeed, due to the two cones the potential wraps both, the NACTs and the propagating beam thus enhancing the interaction in the region close to the ci-point. In other words, altogether, the molecular field gets more and more intense the closer the interaction gets to the ci-point, exactly where the electric beam *funnels* from one PES to the other.

This situation, if it exists, is somewhat reminiscent of what is encountered in astronomy/cosmology (still, to be taken with a grain of salt). Here astronomers considered the possibility that certain stars which are heavy enough may affect due to their extensive mass the path and the nature of a passing-by beam of light. These stars were termed by Wheeler as Black Holes [82] and are characterized by having a well defined region surrounded by what is called "HORIZON" and a singularity point at its center. [82] These two features are somewhat similar to what happens in the present situation. Here the NACTs are located within a spatial region formed by the adiabatic PESs (see Eq. (49)) and found to have, as we know, a singularity, at its center (see Fig. 1). Such a system seems to be capable of affecting (but not necessarily deflecting) the electric beam. However the effect is not produced by a mass, as in the astronomic case, but due to NACTs as is discussed here.

In another article [83] we suggest to move one step further by asking the following: Since we deal with features reminiscent of cosmic phenomena is it likely that Molecular Fields will be enhanced by genuine mass effects? In order to study this possibility the BOH approach was modified by replacing the ordinary NACTs with the mass affected NACTs [83] which are formed by applying the Dirac relativistic theory of the electron. [84-85] It is our intention to replace the ordinary inhomogeneities, $J_p$ and $J_q$ (see Eqs. (45)), of the Wave Equations (see Eqs. (37)) with above mentioned, relativistic mass affected NACTs.

The fact that eventually we are facing mathematics associated with cosmic situations may lead to another question, namely is it possible that the mathematics applied by us to study the (singular) NACTs are related to cosmic phenomena.[86]

However for that to happen we need a situation where <u>at least</u> three nuclei, e.g., three protons are at a close vicinity (up to 10-15 a.u.) thus forming during an instant , e.g., one of the two cations $H_3^+$ or $H_3^{++}$. These NACTs, as well as their virtual singular points, continue to exist even when the inter-atomic distances become, afterwards, large and *eventually* infinite [86] and in this way encode the *formation of molecules*. In other words could it be that the mathematic that treats correctly the Big Bang imitates, in addition to the production of the mass, charge, spin etc. of cosmic particles also the formation of (polyatomic) NACTs and consequently, *guaranty* the creation of molecules during this event? This issue was briefly discussed in Refs. (8b) but more seriously approached in (the forthcoming) Ref. (86).

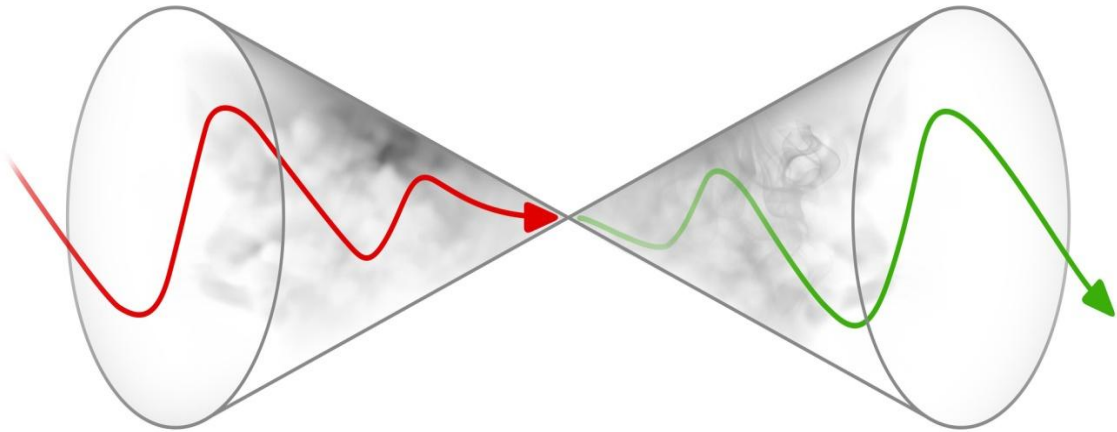

Fig.1. An electric beam, schematically presented by the wiggled red-green arrow, interacting with NACTs (shown in gray) "concentrated" inside the two intersecting conical potentials. Due to the two cones the potentials wrap both, the NACTs and the propagating beam thus enhancing a constructive interaction the closer the beam gets to the ci-point.

# Acknowledgement

I was a Visiting Professor at IISER Mohali, where this work was started (back on 2014). On this occasion I would like to thank Professor N. Sathyamurthy, the Director of IISER Mohali, for organizing my four-month stay in his institute and IISER Mohali of partially supporting it.

# Appendix A

# Field-Free (Stationary) Wave Equation

In what follows is presented the derivation of the *field free*, two-state Wave Equation which is also known as the stationary Wave Equation.. The derivation is based on the fact that both the 2×2 Curl equation and the 2×2 Divergence equation are Abelian, namely these two operators apply also to their off-diagonal matrix elements.

## A.1 Curl Equation

The definition of **F**, namely the Curl for a matrix **τ** is given in the form:

$$\mathbf{F} = \mathbf{H} - \mathbf{T} \qquad (A.1)$$

where the corresponding (p,q) Tensorial matrix element of **H** is:

$$\mathbf{H}_{pq} = (Curl\tau)_{pq} = \frac{\partial \tau_p}{\partial q} - \frac{\partial \tau_q}{\partial p} \qquad (A.2)$$

and the corresponding (p,q) Tensorial matrix element of **T** is:

$$\mathbf{T}_{pq} = [\tau_p, \tau_q] \quad \Rightarrow \quad \mathbf{T} = [\tau \times \tau] \qquad (A.3)$$

Here $[\tau_p, \tau_q]$ is the commutation relation between the matrices $\tau_p$ and $\tau_q$.

Next is mentioned one of the more important features that characterizes this research field: In case of a Hilbert space [3c] (namely, when the applied electronic set of eigen-functions forms, in the given configuration space, a complete basis set) **F** has to be identically zero.[10] Thus:

$$\mathbf{F} \equiv \mathbf{0} \qquad (A.4)$$

We continue by considering the two-state (N=2) system. In this case the 2×2 $\tau_z$-matrix; z=p,q is of the form:

$$\tau_z = \begin{pmatrix} 0 & \tau_{z12} \\ -\tau_{z12} & 0 \end{pmatrix}; \quad z = p, q \qquad (A.5)$$

and consequently $\mathbf{T}_{pq}=0$ and therefore Eqs. (A.2) and (A.4) lead to the following result:

$$\mathbf{H}_{pq} = \begin{pmatrix} 0 & (Curl\boldsymbol{\tau}_{12})_{pq} \\ -(Curl\boldsymbol{\tau}_{12})_{pq} & 0 \end{pmatrix} = 0 \tag{A.6}$$

or

$$(Curl\boldsymbol{\tau}_{12})_{pq} \equiv \frac{\partial \tau_{p12}}{\partial q} - \frac{\partial \tau_{q12}}{\partial p} = 0 \tag{A.7}$$

Eq. (A.7) is the Abelian form of the *Curl* equation in case of a two-state system.

We complete this section with two comments:

(1) The derivations in this section apply as long as the electronic eigen-functions are analytic functions at *every* point in the region of interest. In case certain eigen-functions are not analytic at some points in this region, e.g., at *conical intersections* (*ci*), [5i] the respective $\tau$-matrix element are *singular* and therefore their derivatives are not defined (such points are designated as *pathological* points).

(2) Abelian situations are encountered only at the vicinity of conical intersections. The reason being that close enough to the *ci*-points Hilbert spaces are formed by two states (the two states that form the *ci* )

## A.2. Divergence Equation

Based on Eqs. (5a) and (5b) it was shown [3d] that the Divergence matrix equation for a Hilbert space takes the form:

$$\nabla \boldsymbol{\tau} = \boldsymbol{\tau}^{(2)} - \boldsymbol{\tau}^2 \tag{A.8}$$

where $\boldsymbol{\tau}^{(2)}$ is a matrix that contains the 2-nd order NACTs (see Eq. (5b))

The two-state system is also of special interest for the Divergence equation because here too (just like in case of the Curl equation) the equation for the matrices become an equation for the off-diagonal (1,2) terms. Thus:

$$\nabla \boldsymbol{\tau}_{12} = \frac{\partial \tau_{q12}}{\partial q} + \frac{\partial \tau_{p12}}{\partial p} = \tau_{12}^{(2)} \tag{A.9}$$

where $\tau_{12}^{(2)}$ is introduced in Eq. (4). [3d]

Eq. (A.9) indicates that the Divergence-equation encountered in molecular physics is, in general, different from zero.

## A.3 Wave Equations

Eqs. (A.7) and (A.9) form the *Curl-Div* equations which yield the corresponding, Abelian, Field-free (stationary) Wave Equation..

Thus. differentiating Eq. (A.7) with respect to q and Eq. (A.9) with respect to p and adding them up yields the Wave Equation for $\tau_{p12}$, namely:

$$\frac{\partial^2 \tau_{p12}}{\partial p^2} + \frac{\partial^2 \tau_{p12}}{\partial q^2} = \frac{\partial}{\partial q} \tau_{12}^{(2)} \qquad (A.10)$$

In the same way, differentiating Eq. (A.9) with respect to q and Eq. (A.7) with respect to p and subtracting the second from the first yields the spatial Wave Equation for $\tau_{q12}$, namely:

$$\frac{\partial^2 \tau_{q12}}{\partial q^2} + \frac{\partial^2 \tau_{q12}}{\partial p^2} = \frac{\partial}{\partial p} \tau_{12}^{(2)} \qquad (A/11)$$

Eqs. (A.10) and (A.11) form the field-free Wave Equations for the two (Abelian) components of $\tau_{12}$ [3e,11,12] (they can be extended, straightforwardly, to any number of coordinates). Eqs (A.10) and (A.11) are defined as stationary Wave Equations.

# Appendix B
# Field-Free →Field Dressed Transformation Matrix ω(s,t)

Our study is limited to a system of two (adiabatic) states and therefore all matrices are of 2×2 dimension. Following a rather thorough analysis it was proved that the most general form for the ω(s,t)-matrix, has to be: [7]

$$\boldsymbol{\omega}(\mathbf{s},t) = \begin{pmatrix} e^{i\theta_1} & 0 \\ 0 & e^{-i\theta_2} \end{pmatrix} \begin{pmatrix} \cos\theta/2 & \sin\theta/2 \\ -\sin\theta/2 & \cos\theta/2 \end{pmatrix} \begin{pmatrix} e^{i\varphi} & 0 \\ 0 & e^{-i\varphi} \end{pmatrix} \quad (B.1a)$$

thus, expressed in terms of *four* angles. Within the present study the ω-matrix can be shown to be of the form[1,2,3b]

$$\boldsymbol{\omega}(\mathbf{s},t) = \exp\left\{-\frac{i}{\hbar}\int_0^t \tilde{\mathbf{H}}_e(\mathbf{s},t')dt'\right\} \quad (B.1b)$$

where $\tilde{\mathbf{H}}_e(\mathbf{s},t)$ is the corresponding potential matrix given in the form:

$$\tilde{\mathbf{H}}_e(\mathbf{s},t) = \mathbf{u}(\mathbf{s}) + \tilde{\boldsymbol{\Phi}}_e(\mathbf{s},t) \quad (B.2)$$

Here **u**(**s**) is a diagonal matrix containing the BO adiabatic (TID) potentials and $\tilde{\Phi}_e(\mathbf{s},t)$ is the resulting interaction-matrix due to the external EM field (see Eq. (3) in the main body of the article).

To solve the exponentiated integral in Eq. (B.1) the interval [0,*t*] is divided into n sub-intervals : [0, $t_1$, $t_2$…, $t_k$,…, $t_n$ ] so that Eq.(B.2) becomes:

$$\boldsymbol{\omega}(\mathbf{s},t) = \prod_{k=1}^{n} \exp\left\{-\frac{i}{\hbar} \int_{t_{k-1}}^{t_k} \tilde{\mathbf{H}}_e(\mathbf{s},t')dt'\right\} \sim \prod_{k=1}^{n} \exp\left\{-\frac{i}{\hbar} \tilde{\mathbf{H}}_e(\mathbf{s},\tilde{t}_k)\Delta t\right\} \quad (B.3)$$

where $\tilde{\mathbf{H}}_e(\mathbf{s},\tilde{t}_k)$ is an average value of $\tilde{\mathbf{H}}_e(\mathbf{s},t)$ along the k-th time-interval: {$t_{k-1}$,$t_k$} and $\Delta t$ is the length of the time interval.

Defining $\mathbf{A}_k$ as an orthogonal matrix that diagonalizes $\tilde{\mathbf{H}}_e(\mathbf{s},\tilde{t}_k)$ we get for Eq. (B.3) the following expression:

$$\boldsymbol{\omega}(\mathbf{s},t_n) = \prod_{k=1}^{n} \exp\left\{-\frac{i}{\hbar} \mathbf{A}_k(\mathbf{s},\tilde{t}_k)\boldsymbol{\lambda}_e(\mathbf{s},\tilde{t}_k)\mathbf{A}_k(\mathbf{s},\tilde{t}_k)^{\dagger} \Delta t_k\right\}$$

or

$$\boldsymbol{\omega}(\mathbf{s},t_n) = \prod_{k=1}^{n} \mathbf{A}_k(\mathbf{s},\tilde{t}_k)\exp\left\{-\frac{i}{\hbar} \boldsymbol{\lambda}_e(\mathbf{s},\tilde{t}_k)\Delta t_k\right\}\mathbf{A}_k(\mathbf{s},\tilde{t}_k)^{\dagger} \quad (B.4)$$

where $\boldsymbol{\lambda}_e(\mathbf{s},\tilde{t}_k)$ is a diagonal matrix which contains the eigenvalues of $\tilde{\mathbf{H}}_e(\mathbf{s},\tilde{t}_k)$ and $\tilde{t}_k$ is an intermediate value of $t$ in k-th interval $\{t_{k-1}, t_k\}$. Eq. (B.4) can also be written as:

$$\boldsymbol{\omega}(\mathbf{s},t_n) = \prod_{k=1}^{n} \tilde{\boldsymbol{\omega}}_k(\mathbf{s},\tilde{t}_k) \tag{B.5}$$

where $\tilde{\boldsymbol{\omega}}_k(\mathbf{s},\tilde{t}_k)$ stands for:

$$\tilde{\boldsymbol{\omega}}_k(\mathbf{s},\tilde{t}_k) = \mathbf{A}_k(\mathbf{s},\tilde{t}_k)\exp\left\{-\frac{i}{\hbar}\int_{t_{k-1}}^{t_k}(\boldsymbol{\lambda}_e(\mathbf{s},t)dt)\right\}\mathbf{A}_k(\mathbf{s},\tilde{t}_k)^{\dagger} \tag{B.6}$$

Here $\boldsymbol{\lambda}_e(\mathbf{s},t)$, just like $\boldsymbol{\lambda}_e(\mathbf{s},\tilde{t}_k)$, is a diagonal matrix which contains the eigen-values of $\tilde{\mathbf{H}}_e(\mathbf{s},t)$. Eq. (B.6) concludes the derivation of the $\tilde{\boldsymbol{\omega}}_k(\mathbf{s},\tilde{t}_k)$-matrix along a short time interval $\{t_{k-1}, t_k\}$.

Next we concentrate on the 2-state case for which Eq (B.1) is written, explicitly, as:

$$\tilde{\boldsymbol{\omega}}_k(\mathbf{s},\tilde{t}_k) = \exp\left\{-\frac{i}{\hbar}\int_{t_{k-1}}^{t_k}\begin{pmatrix}\tilde{\mathbf{u}}_1(\mathbf{s},t') & \tilde{\boldsymbol{\Phi}}_{e12}(\mathbf{s},t')\\ \tilde{\boldsymbol{\Phi}}_{e12}(\mathbf{s},t') & \tilde{\mathbf{u}}_2(\mathbf{s},t')\end{pmatrix}dt'\right\} \tag{B.7}$$

where $\tilde{\mathbf{u}}_j(\mathbf{s},t)$ ; j=1,2 are defined, following Eq. (B.2), as:

$$\tilde{\mathbf{u}}_j(\mathbf{s},t) = \mathbf{u}_j(\mathbf{s}) + \tilde{\Phi}_{ejj}(\mathbf{s},t); \quad j = 1,2 \tag{B.8}$$

The eigenvalues of the 2×2 *potential matrix* in Eq. (B.7) under the integral sign (recalling Eq. (B.8)) are:

$$\lambda_e^{\pm}(\mathbf{s},t) = \frac{1}{2}(\tilde{u}_1 + \tilde{u}_2) \pm \frac{1}{2}\sqrt{(\tilde{u}_1 - \tilde{u}_2)^2 + 4\tilde{\Phi}_{e12}^2} \tag{B.9}$$

and the corresponding eigenvectors (that form the orthogonal transformation matrix)

$$\begin{pmatrix} a_1 \\ b_1 \end{pmatrix} = \frac{1}{\sqrt{\tilde{\Phi}_{e12}^2 + (\lambda_e^+ - \tilde{u}_1)^2}} \begin{pmatrix} \tilde{\Phi}_{e12} \\ \lambda_e^+ - \tilde{u}_1 \end{pmatrix} \tag{B.10a}$$

$$\begin{pmatrix} a_2 \\ b_2 \end{pmatrix} = \frac{1}{\sqrt{\tilde{\Phi}_{e12}^2 + (\lambda_e^- - \tilde{u}_2)^2}} \begin{pmatrix} \lambda_e^- - \tilde{u}_2 \\ \tilde{\Phi}_{e12} \end{pmatrix} \tag{B.10b}$$

In what follows we consider the *high intensity* case, namely:

$$\tilde{\Phi}_{e12} \gg \frac{1}{2}|(\tilde{u}_1 - \tilde{u}_2)| \tag{B.11}$$

which guarantees higher transparency regarding the meaning of the results, later to be derived. For this purpose we consider the diagonal matrix, $\tilde{\lambda}_{ek}(\mathbf{s}, t_k)$, in Eq. (B.6):

$$\tilde{\lambda}_{ek}(\mathbf{s}, t_k) = \begin{pmatrix} \exp\left\{-\frac{i}{\hbar}\int_{t_{k-1}}^{t_k}\left(\lambda_e^+(\mathbf{s},t)dt\right)\right\} & 0 \\ 0 & \exp\left\{-\frac{i}{\hbar}\int_{t_{k-1}}^{t_k}\left(\lambda_e^-(\mathbf{s},t)dt\right)\right\} \end{pmatrix}$$

(B.12a)

and introduce the two following phases.

$$\theta_{1k} = -\frac{1}{\hbar}\int_{t_{k-1}}^{t_k}\lambda_e^+(\mathbf{s},t)\mathrm{d}t = -\frac{1}{\hbar}\int_{t_{k-1}}^{t_k}\mathrm{d}t\left[\frac{1}{2}(\tilde{u}_1+\tilde{u}_2)+\tilde{\Phi}_{e12}(\mathbf{s},t)\right]$$

$$\theta_{2k} = \frac{1}{\hbar}\int_{t_{k-1}}^{t_k}\lambda_e^-(\mathbf{s},t)\mathrm{d}t = \frac{1}{\hbar}\int_{t_{k-1}}^{t_k}\mathrm{d}t\left[\frac{1}{2}(\tilde{u}_1+\tilde{u}_2)-\tilde{\Phi}_{e12}(\mathbf{s},t)\right]$$

(B.13)

So that Eq. (B.12a) becomes:

$$\tilde{\lambda}_{ek}(\mathbf{s},t_k) = \begin{pmatrix} \exp i\theta_1(\mathbf{s},t_k) & 0 \\ 0 & \exp -i\theta_2(\mathbf{s},t_k) \end{pmatrix} \quad \text{(B.12b)}$$

Next we consider the corresponding orthogonal transformation matrix, $\mathbf{A}_k(\mathbf{s},\tilde{t}_k)$ which is written in the form:

$$\mathbf{A}_k(\mathbf{s},\tilde{t}_k) = \begin{pmatrix} \cos\theta(\mathbf{s},\tilde{t}_k)/2 & -\sin\theta(\mathbf{s},\tilde{t}_k)/2 \\ \sin\theta(\mathbf{s},\tilde{t}_k)/2 & \cos\theta(\mathbf{s},\tilde{t}_k)/2 \end{pmatrix} \quad \text{(B.14)}$$

where the angle θ is derived employing Eqs. (B.10):

$$\theta(\mathbf{s},\tilde{t}_k) = 2\tan^{-1}\left( \frac{-\lambda_e^-(\mathbf{s},\tilde{t}_k) + \tilde{u}_2(\mathbf{s},\tilde{t}_k)}{\tilde{\Phi}_{e12}(\mathbf{s},\tilde{t}_k)} \right) \quad \text{(B.15)}$$

For the high intensity limit this expression simplifies to become:

$$\theta(\mathbf{s},\tilde{t}_k) = 2\tan^{-1}\left\{ \frac{\tilde{\Phi}_{e12}(\mathbf{s},\tilde{t}_k) + \dfrac{1}{2}\left[\tilde{u}_2(\mathbf{s},\tilde{t}_k) - \tilde{u}_1(\mathbf{s},\tilde{t}_k)\right]}{\tilde{\Phi}_{e12}(\mathbf{s},\tilde{t}_k)} \right\} \quad \text{(B.16)}$$

The next step is to consider the product of two successive $\omega_k$-matrices:

$$\tilde{\omega}_k(\mathbf{s},\tilde{t}_k)\tilde{\omega}_{k+1}(\mathbf{s},\tilde{t}_{k+1}) =$$

$$\mathbf{A}_k(\mathbf{s},\tilde{t}_k)\exp\left\{-\frac{i}{\hbar}\int_{t_{k-1}}^{t_k}\left(\lambda_e(\mathbf{s},t)dt\right)\right\}\times \qquad (B.17)$$

$$\mathbf{J}_{kk+1}(\mathbf{s}\,|\,\tilde{t}_k,\tilde{t}_{k+1})\exp\left\{-\frac{i}{\hbar}\int_{t_k}^{t_{k+1}}\left(\lambda_e(\mathbf{s},t)dt\right)\right\}\mathbf{A}_{k+1}(\mathbf{s},\tilde{t}_{k+1})^{\dagger}$$

where $\mathbf{J}_{kk+1}$ stands for the product:

$$\mathbf{J}_{kk+1}(\mathbf{s}\,|\,\tilde{t}_k,\tilde{t}_{k+1}) = \mathbf{A}_k(\mathbf{s},\tilde{t}_k)^{\dagger}\mathbf{A}_{k+1}(\mathbf{s},\tilde{t}_{k+1}) \qquad (B.18a)$$

At the high intensity limit (see Eq. B.11)) the transformation angle $\theta(\mathbf{s},\tilde{t}_k)$ is only weakly dependent on k (see Eqs. (B.11) and (B.16)) so that the same applies to $\mathbf{A}_k(\mathbf{s},\tilde{t}_k)$ and consequently we have, for any (k,k+1) pair, the following result:

$$\mathbf{J}_{kk+1}(\mathbf{s}\,|\,\tilde{t}_k,\tilde{t}_{k+1}) \sim \mathbf{I} \qquad (B.18b)$$

Next, combining Eqs. (B.5) and (B.6) and recalling Eqs (B.17) and (B.18) we get:

$$\tilde{\omega}(\mathbf{s},t) = \exp\left\{-\frac{i}{\hbar}\int_{t=0}^{t}\left(\lambda_e(\mathbf{s},t)dt\right)\right\}\mathbf{A}(\mathbf{s},t)^{\dagger} \qquad (B.19)$$

(where, without affecting the final results, we may assume that $\mathbf{A}(\mathbf{s}, t=0) = \mathbf{I}$).

Finally, recalling Eqs. (B.12),(B.13), B.(14) and (B.19) we find that the ω-matrix, for the high intensity limit, takes the simplified expression:

$$\boldsymbol{\omega}(\mathbf{s},t) = \begin{pmatrix} \exp i\theta_1(\mathbf{s},t) & 0 \\ 0 & \exp -i\theta_2(\mathbf{s},t) \end{pmatrix} \begin{pmatrix} \cos\theta(\mathbf{s},t)/2 & \sin\theta(\mathbf{s},t)/2 \\ -\sin\theta(\mathbf{s},t)/2 & \cos\theta(\mathbf{s},t)/2 \end{pmatrix}$$
(B.20)

where

$$\theta_1(\mathbf{s},t) = -\frac{1}{\hbar}\int_0^t \left\{ \frac{1}{2}\left[\tilde{u}_1(\mathbf{s},t') + \tilde{u}_2(\mathbf{s},t')\right] + \tilde{\Phi}_{e12}(\mathbf{s},t') \right\} dt'$$

$$\theta_2(\mathbf{s},t) = \frac{1}{\hbar}\int_0^t \left\{ \frac{1}{2}\left[\tilde{u}_1(\mathbf{s},t') + \tilde{u}_2(\mathbf{s},t')\right] - \tilde{\Phi}_{e12}(\mathbf{s},t') \right\} dt'$$
(B.21)

and

$$\theta(\mathbf{s},t) = \tan^{-1}\left\{ \frac{\tilde{\Phi}_{e12}(\mathbf{s},t) - [\tilde{u}_2(\mathbf{s},t) - \tilde{u}_1(\mathbf{s},t)]}{\tilde{\Phi}_{e12}(\mathbf{s},t)} \right\}$$
(B.22)

In addition we also frequently refer to the angle $\Theta(\mathbf{s},t)$ defined as:

$$\Theta(\mathbf{s},t) = \theta_1(\mathbf{s},t) + \theta_2(\mathbf{s},t) = -\frac{2}{\hbar}\int_0^t \tilde{\Phi}_{e12}(\mathbf{s},t')dt' \tag{B.23}$$

It is well noticed that the ω-matrix as derived in Eq. (B.20) and the one presented in Eq. (B.1a) become identical if φ≡0.

We may continue recalling Eq. (24b). In this case the ω-matrix becomes:

$$\boldsymbol{\omega}(\mathbf{s},t) = \frac{\sqrt{2}}{2}\begin{pmatrix} e^{i\theta_1} & e^{i\theta_1} \\ -e^{-i\theta_2} & e^{-i\theta_2} \end{pmatrix}$$

# Appendix C

# Treatment of the Right-Hand-Side of the Wave Equations

In the present Appendix are derived the R.H.S of the Field Dressed Wave Equations derived for the *high intensity* limit (see Eq. (21) in the main body of the article)

Next are introduced the Curl equations for one of the *off-diagonal* matrix elements (reminiscent of how it is done in the Field-Free *Abelian* case – Appendix A). We start by treating the Spatial component of the Field Dressed Curl:

$$C_{pq}(p,q|t) = \frac{\partial \tilde{\tau}_{p12}}{\partial q} - \frac{\partial \tilde{\tau}_{q12}}{\partial p} \tag{C.1a}$$

which due to Eq. (42) becomes (see also Eq. (A.7)):

$$C_{pq} = -\sin\Theta \left( \frac{\partial \Theta}{\partial q} \tau_{p12} - \frac{\partial \Theta}{\partial p} \tau_{q12} \right) \tag{C.1b}$$

Next are treated the two space-time components of the Field-Dressed Curl:

$$C_{pt}(p,q|t) = i\hbar \frac{\partial \tilde{\tau}_{p12}}{\partial t} - \frac{\partial \tilde{\tau}_{t12}}{\partial p} \tag{C.2a}$$

which, due to Eqs. (42) and (43), becomes:

$$C_{pt}(p,q|t) = -i\hbar \sin \Theta(t) \tau_{p12}(p,q) \frac{\partial \Theta}{\partial t} \qquad (C.2b)$$

and

$$C_{qt}(p,q|t) = i\hbar \frac{\partial \tilde{\tau}_{q12}}{\partial t} - \frac{\partial \tilde{\tau}_{t12}}{\partial p} \qquad (C.3a)$$

which, in the same manner, becomes:

$$C_{qt}(p,q|t) = -i\hbar \sin \Theta(t) \tau_{q12}(p,q) \frac{\partial \Theta}{\partial t} \qquad (C.3b)$$

Finally is treated the space-time Field-Dressed Divergence equation:

$$D_{pqt} = \frac{\partial \tilde{\tau}_{p12}}{\partial p} + \frac{\partial \tilde{\tau}_{q12}}{\partial q} + \frac{i}{\hbar c^2} \frac{\partial \tilde{\tau}_{t12}}{\partial t} \qquad (C.4a)$$

which again following the application of Eqs. (42) and (43), yields the required expression for the Divergence

$$D_{pqt}(p,q,t) = \cos\Theta \, \tau_{12}^{(2)}(p,q) -$$

$$\sin\Theta \left( \tau_{p12}(p,q)\frac{\partial\Theta}{\partial p} + \tau_{q12}(p,q)\frac{\partial\Theta}{\partial q} \right) + \qquad (C.4b)$$

$$\frac{i}{2}\left( \frac{\partial^2\Theta}{\partial p^2} + \frac{\partial^2\Theta}{\partial q^2} - \frac{1}{c^2}\frac{\partial^2\Theta}{\partial t^2} \right)$$

Next is introduced one additional approximation, namely, the duration of the electric perturbation is assumed to be short enough for the nuclei barely to move. To validate this approximation all derivatives of the kind

$$\frac{\partial^n\Theta}{\partial z^n} = 0; \quad z=p,q; \quad n=1,2 \qquad (C.5)$$

have to be incorporated in Eqs. (C.1b) and (C.4b) (the two other equations are not affected). As a result Eq. (C.1b) becomes:

$$C_{pq} = 0 \qquad (C.1b')$$

and Eq. (C.4b) becomes:

$$D_{pqt} = \cos\Theta \, \tau_{12}^{(2)}(p,q) - \frac{i}{2}\frac{1}{c^2}\frac{\partial^2\Theta}{\partial t^2} \qquad (C.4b')$$